# Prospect for antiferromagnetic spintronics

X. Martí[1,2,*], I. Fina[3,4,†], T. Jungwirth[1,5]

[1] Institute of Physics ASCR, v.v.i., Cukrovarnicka 10, 162 53 Praha 6, Czech Republic
[2] Centre d'Investigacions en Nanociencia i Nanotechnologia (CIN2), CSIC-ICN, Bellaterra 08193, Barcelona, Spain
[3] Max Planck Institute of Microstructure Physics, Weinberg 2, Halle (Saale), D-06120, Germany
[4] Department of Physics, University of Warwick, Coventry CV 4 7AL, United Kingdom
[5] School of Physics and Astronomy, University of Nottingham, Nottingham NG7 2RD, United Kingdom

**Exploiting both spin and charge of the electron in electronic micordevices has lead to a tremendous progress in both basic condensed-matter research and microelectronic applications, resulting in the modern field of spintronics. Current spintronics relies primarily on ferromagnets while antiferromagnets have traditionally played only a supporting role. Recently, antiferromagnets have been revisited as potential candidates for the key active elements in spintronic devices. In this paper we review approaches that have been employed for reading, writing, and storing information in antiferromagnets.**

*Keywords*—spintronics, antiferromagnets

## I. INTRODUCTION

ELECTRICALLY erasable programmable read-only memory and flash memory, or ferroelectric random-access memory are examples of non-volatile technologies based on charge storage.[1] Several emerging concepts, including the crystalline-amorphous-chalcogenide phase change memory, insulating-conductive-oxide memristor, or ferromagnetic magnetoresistive random-access memory (MRAM) are based on resistance switching.[1,2] MRAM is unique among all these devices in that it utilizes electron spin for its functionality.

Compared to the above broad family of charge-based memories, all commercial spin-based devices rely on one principle in which the opposite magnetic moment orientations in a ferromagnet (FM) represent the "zeros" and "ones" (see Fig. 1a). This technology is behind memory applications ranging from kilobyte magnetic stripe cards to megabyte MRAMs, and terabyte computer hard disks. Since based on spin, the devices are robust against charge perturbations, i.e., are intrinsically radiation-hard. Moreover, they are non-volatile and, compared to flash memory, MRAM offers short read/write times suitable for main random-access computer memories.

MRAM can be viewed as a solid-state-memory variant of the hard disk in which the magnetic medium for storing and the magnetoresistive read-element are merged into one piece of a FM. Unlike hard disk, the magnetic stray field is not anymore needed for reading the state of the bit and the most advanced spin-transfer-torque MRAMs do not even use electromagnets coupled to the ferromagnetic moment of the bit for writing.[2,3] Still the tradition of using FMs, set historically by the hard disk and magnetic tape media, continues to date in the main-stream research and development of magnetic spin-based devices.

Since magnetic fields are not ubiquitous anymore in reading and writing in advanced spintronics, it appears natural to step out of the narrow box limited by FMs and consider also antiferromagnets (AFMs) in which magnetic order is accompanied by a zero net magnetic moment. Seminal work on theory and experiments covered by MacDonald and coworkers mainly focused on AFM counterparts to the GMR,[4,5,6,7] and it was extended by other works in which the stringent requirements of GMR were circumvented.[8,9,10] AFM materials are magnetic inside, however, their microscopic magnetic moments sitting on individual atoms alternate between two opposite orientations (Fig. 1c). This antiparallel moment configuration makes the magnetism in AFMs invisible on the outside. While FM moments can be unintentionally reoriented and the memory erased by disturbing magnetic fields (Fig. 1b), AFMs are not perturbed by even strong magnetic fields (Fig. 1d).

Put in a broader perspective, AFMs can share with FMs the intrinsic radiation-hardness, non-volatility, and speed, which are the important merits making magnetic memories attractive complements to charge-based devices. In addition, however, AFM can bring to magnetic memories insensitivity to strong magnetic field perturbations, whether generated externally or internally within the memory circuitry, are readily compatible with metal or semiconductor electronic structure,[9,11,12] and can offer ultra-fast writing schemes unparalleled in FMs.[13,14]

On equal footing, it has to be also recalled that the insensitivity to magnetic fields and the lack of magnetic stray fields make AFMs significantly more challenging than FMs to explore and understand their behavior, and to exploit these magnets with hidden order in applications. The attractions and challenges are, hand-in-hand, driving the emerging field of AFM spintronics.

## II. METHODOLOGIES FOR SENSING AND MANIPULATING ANTIFERROMAGNETIC MOMENTS

A physical foundation for a class of phenomena that can be used to electrically detect the AFM moments was presented in Ref. 8. It was realized that conceptually the relativistic anisoropic magnetoresistance (AMR) phenomena are equally present in AFMs as in FMs. Since AMR is an even function of the microscopic magnetic moment vector it is the direction of the spin-axis rather than the direction of the macroscopic magnetization that primarily determines the effect. In collinear FMs the two directions are equivalent.

---

[*] xavi.marti@igsresearch.com
[†] ignasifinamartinez@gmail.com

For the Néel-order spin configuration of compensated AFMs only the spin-axis can be defined while the macroscopic magnetization is zero.

Theoretically there is a qualitative difference between the microscopic origins of the ohmic non-crystalline and crystalline AMR components.[16] Since the former component depends only on the angle between the magnetization **m** and current **I**, the effects of the rotating magnetization on the equilibrium electronic structure do not contribute to the non-crystalline AMR. Instead, in the leading order, the non-crystalline AMR reflects the difference between transport scattering matrix elements of electrons with momentum parallel to the current in the **I**//**m** and **I**⊥**m** configurations.[16,17] The difference between scattering matrix elements for an electron with the momentum parallel and perpendicular to the spin axis is due to the relativistic spin-orbit coupling.

In the transport geometry corresponding to the pure crystalline AMR, magnetization rotates in the plane perpendicular to the current so that the angle between magnetization and current remains constant and the AMR is determined solely by the varying angle between magnetization and crystal axes. Unlike the non-crystalline AMR, the crystalline AMR originates from the changes in the equilibrium relativistic electronic structure induced by the rotating magnetization. The picture applies not only to the ohmic regime[12,16] but also to AMR-like effects which were more recently discovered in tunneling devices.[10,19]

A ~1% non-crystalline ohmic AMR was used to electrically detect two distinct magnetic states in an FeRh AFM,[17] in a complete analogy to the ~1% AMR of NiFeCo based bits in the first generation of FM MRAMs.[18] A crystalline ohmic AMR in $Sr_2IrO_4$ was used to demonstrate the feasibility of the AFM spintronics in semiconductors.[12]

For short read-out times and small bit dimensions, larger magnetoresistive switching signals are required in current commercial MRAMs with FM tunnel junction bits.[2] Again in analogy with the development of FM spintronics, very large (~100%) magnetoresistance switching signals were detected in AFM tunnel devices due to the tunneling AMR.[10] In both the ohmic and tunneling AFM-AMR devices a bistability at zero magnetic field, i.e. a memory functionality, was observed in the AFM and the insensitivity of the memory states was tested to fields as high as 9 T.[16]

The AFM spin-axis direction can be controlled by magnetic field via an attached exchange-coupled FM[10,12] or, without the auxiliary FM, by technique analogous to heat-assisted magnetic recording.[17,20] An illustration of the latter technique is discussed in detail in the following section. As with heat assisted FM-MRAM devices,[20,21] the time scales and energy efficiency of this method for controlling the AFM spin-axis direction are limited. To circumvent this, a novel mechanism has been predicted that does not involve exchange-coupled FM, heating, or magnetic field. In analogy to current induced torques in FMs, an electrical current is suggested to generate non-equilibrium Néel-order fields exerting opposite torques on the two spin sublattices.[14] The effect might yield ultra-short times for AFM spin-axis reorientation. The AFM dynamics is typically 2-3 orders of magnitude faster than in FMs and can reach ~ps.[13] In AFMs, the spin-axis reorientation due to current-induced torques will then be limited only by the circuitry time-scales for delivering the electrical pulses which can be of order ~10's-100's ps.[22]

Concerning the manipulation of AFM thin layers using the exchange-coupling to an adjacent FM material, there are important limitations for the relative thicknesses ($t_{AFM}/t_{FM}$) of AFM/FM layers that must be considered carefully.[24,25] The optimal manipulation of the staggered AFM moments is obtained when the condition $t_{AFM}/t_{FM} \ll 1$ is fulfilled. In this scenario, minimization of the anisotropy energy[25] implies that AFM moments are reoriented by FM moments which align with the applied external magnetic field as long as the field is stronger than the FM coercive field. In contrast, if $t_{AFM}/t_{FM} \gg 1$, the overall magnetic anisotropy energy is dominated by the thicker AFM layer anisotropy. Since the AFM moments are more rigid to external magnetic fields, the FM layer will be pinned by the AFM layer hindering a possibility to manipulate the AFM moments. Therefore, the tuning of $t_{AFM}/t_{FM}$ is crucial but the precise thickness to be employed must be optimized for each particular set of AFM/FM bilayers. In this direction, to decide if the thicknesses of the layers are optimal, the two scenarios described can be discriminated by magnetometry measurements. The coercive field ($H_c$) and the exchange bias ($H_{eb}$) are critical indicators that can discriminate among both scenarios and can be obtained by collecting magnetization loops of the bilayers. The two parameters ($H_c$ and $H_{eb}$) can be inferred from the features of the magnetization loops: $H_{eb}$ is the horizontal displacement the loop with respect to zero magnetic field and $2*H_c$ is the width of the loop. $H_c \sim 0$ and $|H_{eb}| > 0$ are typical for $t_{AFM}/t_{FM} \gg 1$ (when the FM is pinned by the AFM), while $H_c > 0$ and $|H_{eb}| \sim 0$ correspond to the case of $t_{AFM}/t_{FM} \ll 1$ (when AFM moments rotate with FM moments). Of particular interest is the intermediate case ($t_{AFM}/t_{FM} \sim 1$) in which both effects are competing. The combination of both pinning of the FM by the AFM and dragging of the AFM by the FM is reflected in magnetization loops comprising both the horizontal shift and the broadening. This intermediate scenario can lead to a partial and hysteretic tilt of AFM moments upon the full reversal of the FM. As a result, biastable states can be realized at zero magnetic field with a different AFM moment direction and, therefore, showing a distinct AMR signal. This functionality was illustrated in the work by Park et al. using IrMn/Py bilayers[10,25]. The work also highlighted that the optimal thicknesses can depend on temperature which has to be accounted for when optimizing the AFM/FM bilayer structure.

### III. SAMPLE PREPARATION

In this and the following section we focus on FeRh used in the AFM-AMR based memories utilizing the heat-assisted magnetic recording. Thin films were prepared by d.c. sputtering. The MgO(001) substrate was first heated to 800 K for 2 hours in $10^{-8}$ torr and then the temperature was lowered to 550 K. Subsequently, Ar gas was introduced (3 mtorr) and the films were grown using 50 W power at a rate of 1 nm min$^{-1}$ using a stoichiometric FeRh target. The plasma was placed at the very top of the substrate and samples were spinning at 5 Hz during the deposition. Samples were cooled down at a rate of 100 K min$^{-1}$ until room temperature. 1 nm Ta capping was deposited to protect the surface of the layer. Structural and magnetic characterization is reported in Ref. 17. Briefly, stoichiometry analyses revealed a small Fe deficiency (Fe/Rh ~ 0.95) for the films, and Fe/Rh ~ 1.00 for the sputtering target. X-ray reciprocal space maps revealed that the samples are epitaxial with both in-plane and out-of-plane textures. The samples show the (001) out-of-plane texture with no traces of spurious phases and orientations. Sharp and hysteretic magnetic and electric resistance transitions were observed with negligible traces of spurious magnetic moment in the room-temperatute AFM phase of FeRh.

## IV. RESULTS

We now illustrate the AFM memory concept on a FeRh ohmic resistor device. Reading of the information is via the AMR and writing by a variant of the heat-assisted magnetic recording technique. FeRh undergoes an AFM to FM transition[16,20,25] at ~100 °C which can by utilized for writing the memory state.[17] In Fig. 2 we show the result of the field-cooling from the FM to the AFM state. Two traces are displayed corresponding to a parallel (blue curve) or perpendicular (red curve) direction of the applied magnetic field during cooling with respect to the direction of the detection current. The subsequent AMR measurement was performed at zero magnetic field while increasing the temperature from 300 to 400 K. We plot the resistances relative to the resistance achieved after cooling the sample in a field oriented perpendicular to the current. Clearly, two distinct resistance states are written in the AFM by the two orthogonal directions of the magnetic field applied during field-cooling. (Note that the AMR peak at ~370 K corresponds to the AFM-FM magnetic phase transition.)

In Fig. 3 we show a result of several consecutive writing cycles with the same AMR definition as in Fig. 2. The data points were recorded at room temperature and zero magnetic field. The „1,0,1,0" sequence written by the field-cooling is read-out by the room-temperature AMR of the AFM memory. Therefore, without prior information of the field-cooling direction, the room-temperature resistance of the FeRh bit allows us to detect its stable memory state. While staying in the AFM phase the memory cannot be erased by strong magnetic fields.

## V. DISCUSSION OF ALTERNATIVE MATERIAL SYSTEMS

Materials displaying the paramagnet-to-FM-to-AFM sequence of magnetic transitions close to room temperature are fairly scarce. Another material system which shows a similar magnetic phase diagram is Cr-doped $Mn_2Sb$. While $Mn_2Sb$ is a ferrimagnet (Tc ~ 550K),[27] x~0.1 chemical substitution of Mn by Cr in $Mn_{2-x}Cr_xSb$ triggers the occurrence of a low-temperature AFM ground state. Close to the x ~ 0.2, the AFM transition temperature crosses room temperature.[27] $Mn_{2-x}Cr_xSb$ is tetragonal (a ~ 4.07 Å and c ~ 6.54 Å) and, similar to FeRh, its magnetic moments are in the plane and along the main crystallographic directions. Therefore, $Mn_{2-x}Cr_xSb$ is another promising material for the heat-assisted magnetic recording in an AFM which circumvents the need for expensive Rh. The $Mn_2Sb$ family is particularly interesting as chemical substitutions of Mn by transition metals can lead to a plethora of scenarios of high interest in the field of AFM spintronics. For instance, increasing doping with Co triggers a ferrimagnet-to-AFM phase transition and, akin to Cr, enables an intermediate FM phase. However, beyond x~0.35 in $Mn_{2-x}Co_xSb$, the FM intermediate phase vanishes and the system transits directly from the AFM to the PM phase. Therefore, a sequence of Co-doped $Mn_2Sb$ samples could allow to compare the field-cooling procedure with and without the FM intermediate phase. Such experiments would allow concluding how critical the intermediate FM phase is in the arrangement of the magnetic moments in the AFM phase during a field-cooling process. In this context we recall that field-cooling across the IrMn paramanget-to-AFM transition in polycrystalline thin films[20] allowed to set two distinct metastable tunnel-resistance states. The magnitude of the magnetic field applied and the role of the relative field-cooling and magnetic moment directions are still to be understood in more detail. Here the relatively simple tetragonal structure of Co-doped $Mn_2Sb$ appears as a promising test-bed material to contribute to the understanding of the field-cooling process.

Finally, Cr is a spin density wave AFM with a paramagnet-to-AFM transition close to room temperature.[28] Alloying with small amounts of Pt increases $T_N$ of pure Cr (311.5 K) to 440 K (at 0.6 % of Pt), 563 K (at 2 % of Pt), and 460 K (at 5 % of Pt).[29] Similarly, $T_N$ increases linearly with increasing Sn concentrations up to 394 K at 1.4 % of Sn. (Presumably, this composition corresponds to the solubility limit of the Cr-Sn solid solution.[29]) Manipulation of Cr AFM domains by field-cooling has already been explored.[28] Single domain AFM states can be set by applying relatively small magnetic fields when cooling across $T_N$. The material may be therefore promising for the realizing the concept of heat-assisted magnetic recording in AFMs.

## VI. SUMMARY

To conclude, we have reviewed several basic principles for reading, writing, and storing information in AFMs. We have highlighted potential advantages in using AFMs instead of FMs as active components in spintronic devices, and recalled challenges in working with AFMs. Examples of experimental AFM memory devices realized to date show the viability of the emerging field of AFM spintronics.

*We acknowledge support from the ERC Advanced grant no. 268066, from the Ministry of Education of the Czech Republic grant no. LM2011026, from the Grant Agency of the Czech Republic grant no. 14-37427G, from the Academy of Sciences of the Czech Republic Praemium Academiae. I.F. acknowledges the Beatriu de Pinós postdoctoral scholarship (2011 BP-A 00220) from the Catalan Agency for Management of University and Research Grants (AGAUR-Generalitat de Catalunya).*

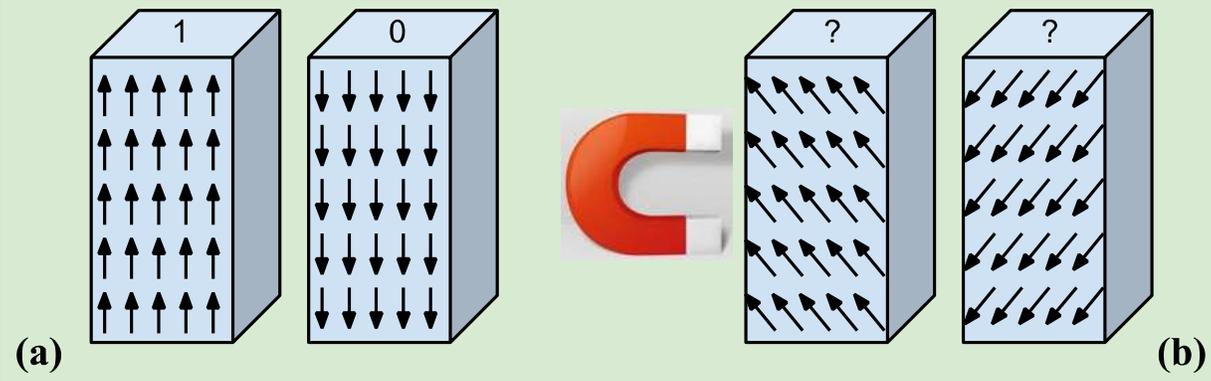
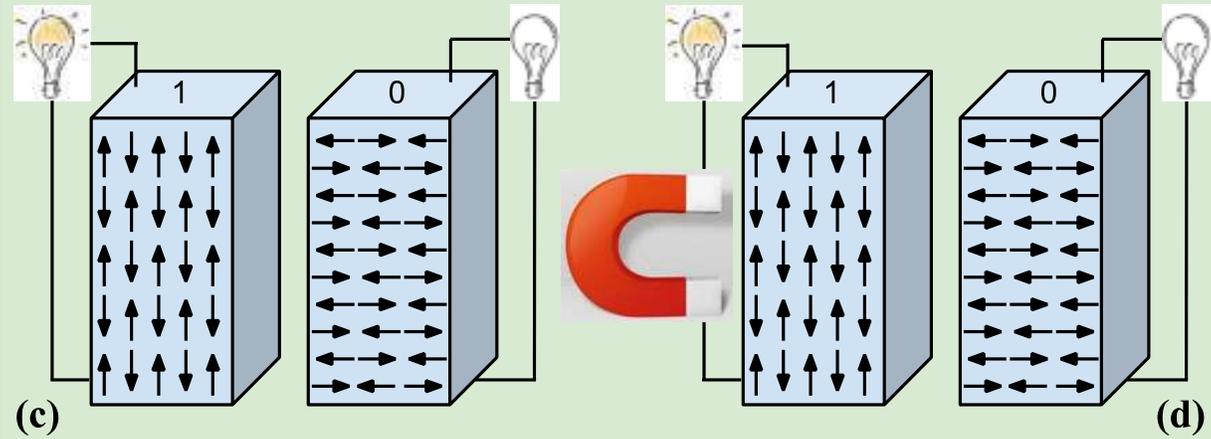
**Figure 1.** a,b FM memory sensitive to magnetic field perturbations. c,d AFM memory insensitive to magnetic field perturbations.

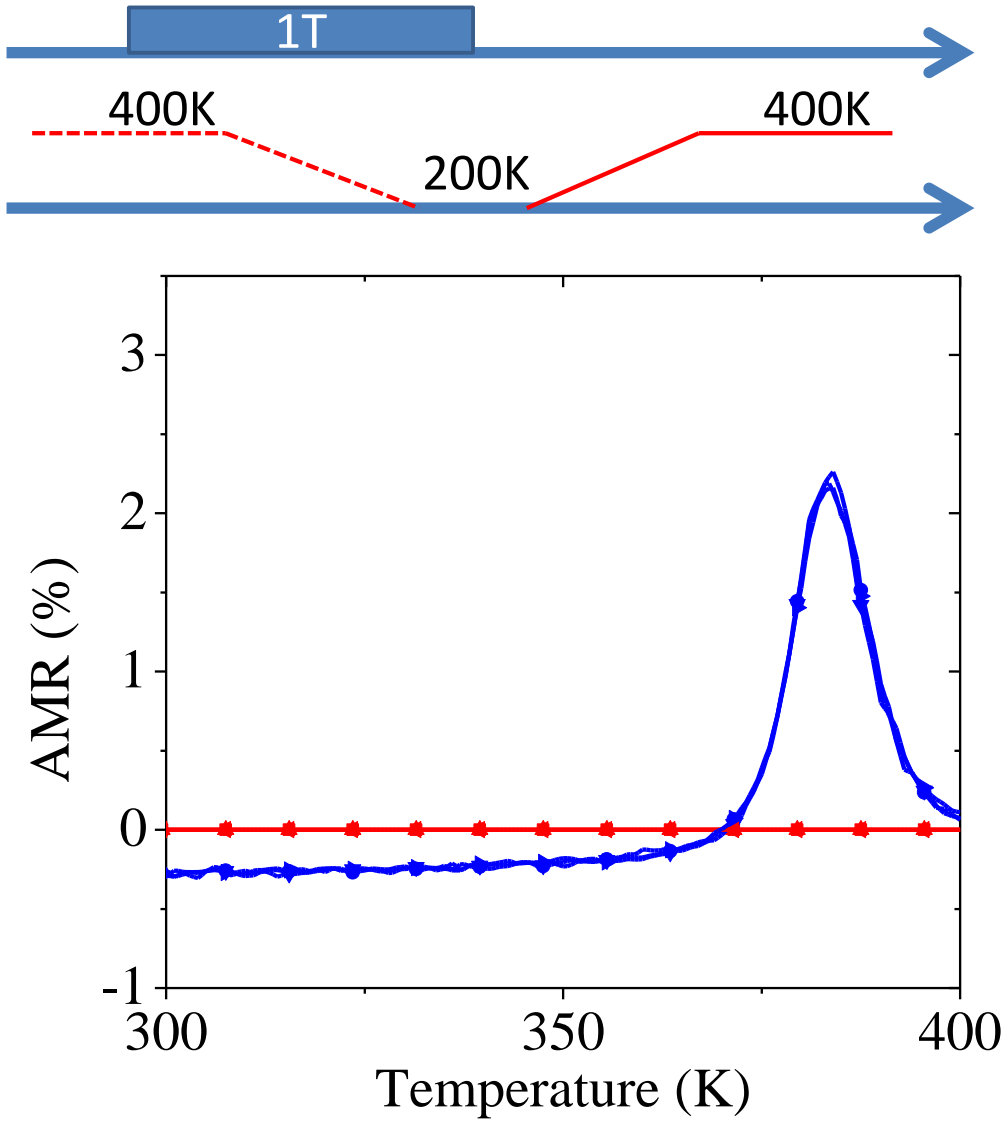

**Figure 2** AMR dependence on temperature while heating the sample at a zero magnetic field. AMR(T) = (R$_{\perp,\parallel}$(T) - R$_{\perp}$(T))/R$_{\perp}$(T). Prior to the AMR measurements the sample was first brought safely above the AFM-FM transition and then cooled back to the AFM state at a magnetic field of 1 T applied either parallel ($\parallel$, blue trace) or perpendicular ($\perp$, red trace) to the current direction in the AMR measurement. The sequence of measurement steps is highlighted in the top schematic diagram.

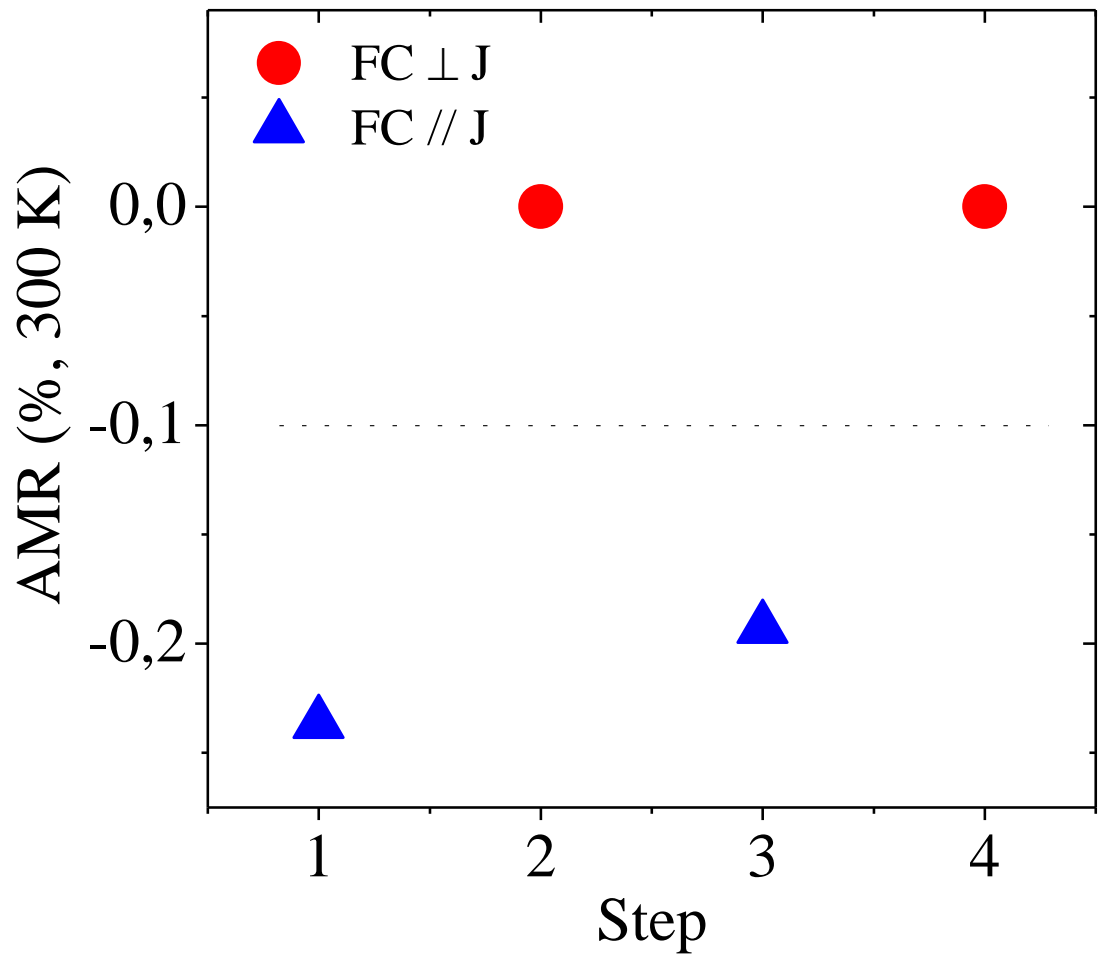

**Figure** 3 Two memory states detected by the AFM-AMR at room temperature after field cooling the sample in a field parallel or perpendicular to the detection current direction.